\documentclass[
    ,final            
  ]
  {aipproc}

\layoutstyle{8x11double}

\begin{document}

\title{The proton microquasar}


\author{Gabriela S. Vila}{
  address={Instituto Argentino de Radioastronom\'ia (CCT La Plata - CONICET)\\
 				 C.C. 5 (1894), Villa Elisa, Buenos Aires, Argentina}
}

\author{Gustavo E. Romero}{
  address={Instituto Argentino de Radioastronom\'ia (CCT La Plata - CONICET)\\
 				 C.C. 5 (1894), Villa Elisa, Buenos Aires, Argentina}
  ,altaddress={Facultad de Ciencias Astron\'omicas y Geof\'isicas, UNLP\\
  Paseo del Bosque s/n  (1900), La Plata, Buenos Aires, Argentina} 
}

\begin{abstract}
We present a model for high-energy emission in microquasars where the energy content of the jets is dominated by relativistic protons. We also include a primary leptonic component. Particles are accelerated up to relativistic energies in a  compact region located near the base of the jet, where most of the emission is produced. \\
We calculate the production spectrum due to proton and electron synchrotron radiation and photohadronic interactions. The target field for proton-photon collisions is provided by the synchrotron radiation in the acceleration region. In models with a significant leptonic component, strong internal photon-photon absorption can attenuate the emission spectrum at high energies.\\
Depending on the values of the parameters, our model predicts luminosities in the range $10^{34}-10^{37}$ erg s$^{-1}$ up to GeV energies, with a high-energy tail that can extend up to $10^{16}$ eV. In some cases, however, absorption effects can completely suppress the emission above 10 GeV, giving rise to different spectral shapes. These results can be tested in the near future by observations with instruments like GLAST-Fermi, HESS II and MAGIC II.
\end{abstract}

\maketitle

\section{Introduction}

Microquasars are binary systems formed by a donor star that feeds a compact object (black hole or neutron star), where part of the accreted matter is ejected from the system as two collimated non-thermal jets. Depending on the mass of the donor star, microquasars are classified into high-mass or low-mass systems. The emission spectrum of microquasars covers almost the entire electromagnetic spectrum, from radio wavelengths to hard X-rays. Three high-mass X-ray binaries have also been detected at Tev gamma rays gamma-rays. The gamma rays can be produced by interaction of relativistic particles in the jet with the radiation field and the matter in the winds of the companion star. Several models to explain the gamma-ray emission in high-mass microquasars can be found in the literature, see for example \cite{Romero2005}, \cite{BR2006} and \cite{BR2007}. However, in the case of low-mass microquasars, the density of the radiation and matter fields supplied by the donor star is much lower, and the same type of mechanisms proposed for high-mass systems turns out to be inefficient \cite{Grenier2005}. Here we present a model for the high-energy emission of low-mass microquasars, based on the interaction of relativistic particles in the jets with the internal radiation, matter and magnetic field of the jet itself. The model can be also applied to high-mass microquasars, taking into account the effect of the external fields. In our model the high-energy emission is mostly due to hadronic interactions, though we also take into account the contribution of primary and secondary electrons. The results show that, under certain physical conditions, low-mass microquasars can be sources of gamma rays detectable by the satellite GLAST-Fermi, or Cherenkov telescope arrays like HESS II and MAGIC II.

\section{Model description}

\begin{figure}[!h]
  \includegraphics[trim=90 570 280 80,clip,height=.15\textheight]{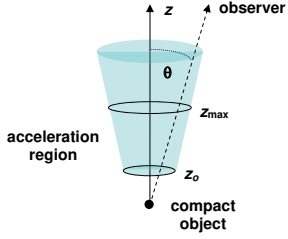}
  \caption{A detail of the jet injection region. The relevant geometrical parameters are indicated.}
  \label{fig:acc_region}
\end{figure}

We consider a conical jet, perpendicular to the orbital plane of the binary. A scheme of the jet launching region is shown in Figure \ref{fig:acc_region}. The jet is launched at a distance $z_0$ from the compact object and has an initial radius $r_0=0.1z_0$. The outflow is assumed to be only mildly relativistic, with a bulk Lorentz factor $\Gamma_{\rm{jet}} = 1.5$. It carries a fraction of the accretion power, $L_{\rm{jet}} = 0.1L_{\rm{accr}}\approx 10^{38}$ erg s$^{-1}$. We further assume that a fraction $L_{\rm{rel}} = 0.1L_{\rm{jet}}$ of the jet kinetic power is in the form of relativistic protons and leptons, $L_{\rm{rel}} = L_p+L_e$. We relate the injected power in protons and electrons through the parameter $a$, $L_p = aL_e$.\\

Particles are accelerated by diffusive shock acceleration leading to a power-law injection, 

\begin{equation}
Q =Q_0\frac{ E^{-\alpha}}{z}\quad\left[\rm{erg}^{-1} \rm{cm}^{-3} \rm{s}^{-1}\right].
\label{injection}
\end{equation}

\vspace{0.2cm}

The acceleration region is compact, extending from the base of the jet up to $z_{\rm{max}} = 5\,z_0$. The efficiency of the acceleration mechanism is characterized by the parameter $\eta$, so that the rate of energy gain for a particle of energy $E$ is 

\begin{equation}
t_{\rm{acc}}^{-1} = \eta ecB/E.
\label{acc_rate}
\end{equation}

\vspace{0.2cm}

\noindent Here we assume an efficient accelerator with $\eta = 0.1$.\\
The magnetic field in the jet decreases as 

\begin{equation}
B =B_0\frac{z_0}{z},
\label{B_field}
\end{equation}

\vspace{0.2cm}

\noindent We determine $B_0=B(z_0)$ by requiring equipartition between magnetic and kinetic energy densities,

\begin{equation}
\frac{B^2_0}{8\pi} = \frac{L_{\rm{jet}}}{2\pi r_0^2 v_{\rm{jet}}},
\label{equipartition}
\end{equation}

\begin{figure}[!h]
\begin{tabular}{cc}
\includegraphics[trim=10 18 10 0,clip,height=.25\textheight]{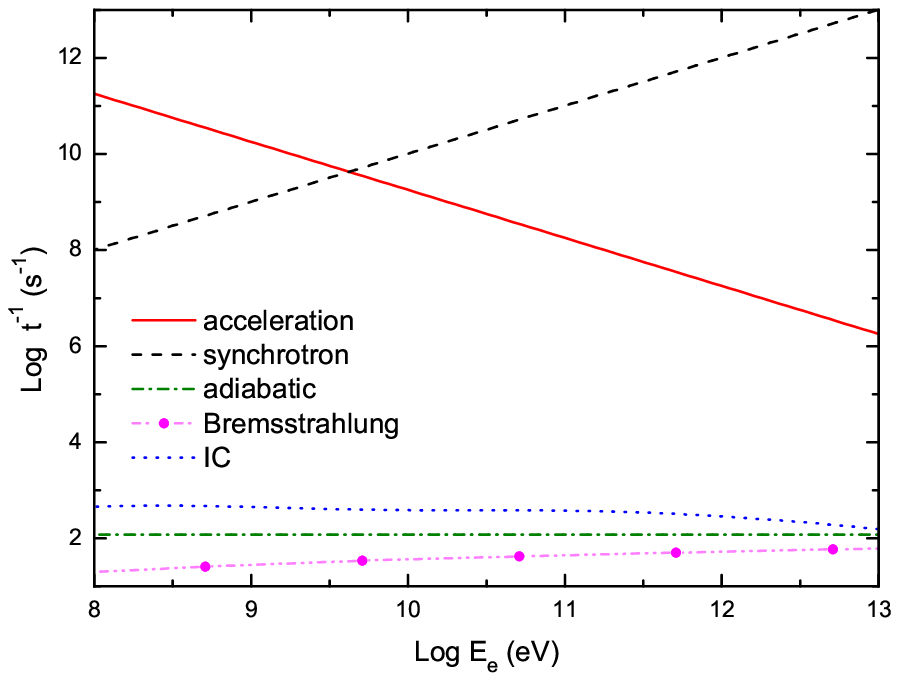} & \includegraphics[trim=10 18 10 0,clip,height=.25\textheight]{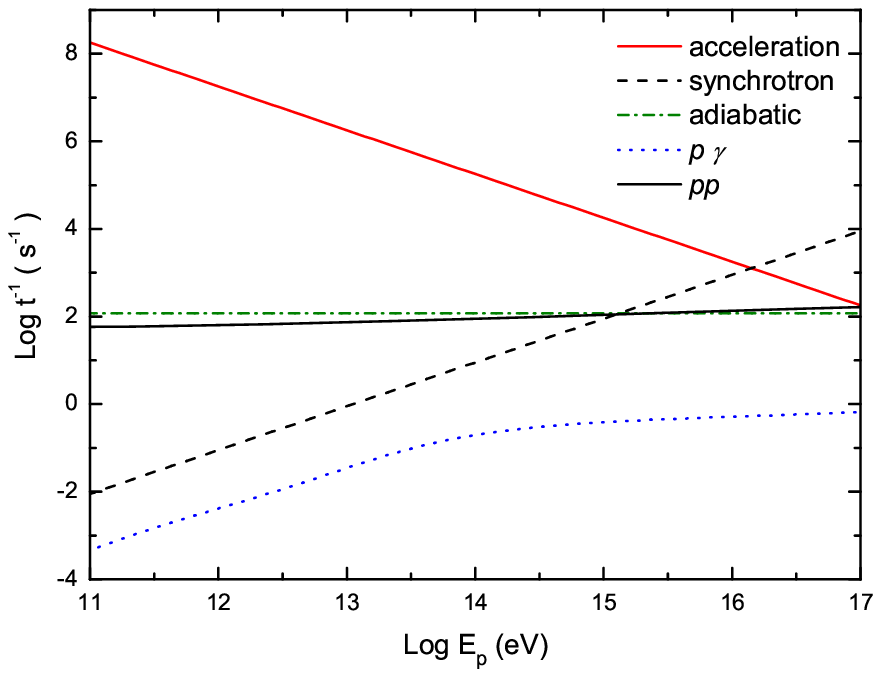} \\
\end{tabular}
  \caption{Acceleration and cooling rates for electrons (left panel) and protons (right panel), calculated at $z_0$ for $a=1$.}
  \label{fig:cooling}
\end{figure}

\vspace{0.2cm}

\noindent where $v_{\rm{jet}}$ is the jet bulk velocity. Equation (\ref{equipartition}) yields $B_0\approx10^7$ G .\\
Figure \ref{fig:cooling} shows the acceleration rate $t_{\rm{acc}}^{-1}$ and the cooling rates $t_{\rm{cool}}^{-1}=E^{-1}dE/dt$ for the different processes of energy loss. The only relevant cooling channel for leptons is synchrotron radiation. This is also true for high-energy protons, whereas at low energies proton cooling is dominated by inelastic proton-proton ($pp$) collisions and adiabatic losses. The maximum energy of the particles is fixed equating acceleration rate and the sum of the cooling rates.  This condition yields $E_{\rm{max}, p}\approx10^{16}$ eV and $E_{\rm{max}, e}\approx10^{10}$ eV. 

Particle distributions  in  steady  state  $N(E,z)$ (in units of erg$^{-1}$ cm$^{-3}$) are calculated  solving the transport equation in the one-zone approximation \cite{Khang2007},

\begin{equation}
\frac{\partial}{\partial E}\left[\frac{dE}{dt}N\right]+\frac{N}{T_{\rm{esc}}}=Q.
\label{transport_eq}
\end{equation}

\vspace{0.2cm}

We consider several photon production mechanisms: proton and electron synchrotron radiation, electron relativistic Bremsstrahlung, proton-proton $(pp)$ inelastic collisions, inverse Compton scattering (IC) and proton-photon $(p\gamma)$ collisions on the synchrotron radiation fields of both protons and electrons. Both $pp$ and $p\gamma$ interactions create neutral $\pi$-mesons that then decay to give gamma-rays,

\begin{equation}
\pi^0\rightarrow2\gamma.  
\label{neutral_pion_decay}
\end{equation}

\vspace{0.2cm}

\noindent To estimate the spectrum from the $\pi^0-$decay we followed Refs. \cite{Kelner2006} and \cite{AD2003} for the cases of $pp$ and $p\gamma$ collisions, 
respectively (for more accurate expressions in the latter case see \cite{KA2008}). Proton-photon and proton-proton collisions also inject high-energy secondary electron-positron pairs, product of the decay of charged pions. Pairs are also injected directly through photopair production,

\begin{equation}
p + \gamma\rightarrow p + e^- + e^+.  
\label{neutral_pion_decay}
\end{equation}

\vspace{0.2cm}

\noindent These secondary leptons also contribute to the gamma-ray emission through synchrotron radiation. To calculate the spectrum of pairs, we used the formulae given by Refs. \cite{Ch1992} and \cite{Mast2005}. The IC spectra were calculated in the local approximation of Ref. \cite{Ghis1985}, whereas for synchrotron and Bremsstrahlung radiation we used classical expressions, see for example Ref. \cite{BG1970}.\\  
All  calculations, except those of interactions with matter (for these see \cite{Reynoso2008}), were performed in the jet co-moving reference frame and the results were transformed to the observer frame using the appropriate Doppler factor $D(\theta)=\left[\Gamma_{\rm{jet}}\left(1-\beta_{\rm{jet}}\cos\theta\right)\right]^{-1}$. We fixed $\theta=30^\circ$ for the viewing angle. See Table \ref{tab:a} for detailed values of the model parameters. 

\begin{table}[!h]
\begin{tabular}{ll}
\hline
   \tablehead{1}{l}{b}{Parameter} & \tablehead{1}{l}{b}{Value}\\
\hline
Jet injection point & $z_0=10^8$ cm\tablenote{$50\,R_{\rm{Schw}}$ for a $8\,M_\odot$ black hole.} \\
Jet initial radius & $r_0=0.1\,z_0$ cm \\
Size of acceleration region & $z_{\rm{max}}=5\,z_0$ \\
Jet bulk Lorentz factor & $\Gamma_{\rm{jet}}=1.5$ \\
Viewing angle & $\theta=30^\circ$ \\
Jet kinetic power & $L_{\rm{jet}}=1.7\times 10^{38}$ erg s$^{-1}$ \\
Proton-to-lepton energy ratio & $a=1-1000$ \\
Acceleration efficiency & $\eta=0.1$ \\
Magnetic field at $z_0$ & $B_0=2\times 10^7$ G \\
Minimum proton/electron energy & $E_{\rm{min}(p,e)}=2-100\,m_{(p,e)}c^2$ \\
Maximum proton/electron energy & $E_{\rm{max}(p,e)}\approx 3\times 10^{16}/10^{10}$ eV\tablenote{Maximum values attained along the jet.} \\
\hline
\end{tabular}
\caption{Values and ranges considered for the various parameters characterizing the jet and the distributions of relativistic particles.}
\label{tab:a}
\end{table}

\begin{figure}[!h]
\begin{tabular}{cc}
\includegraphics[trim=10 18 10 0,clip,height=.25\textheight]{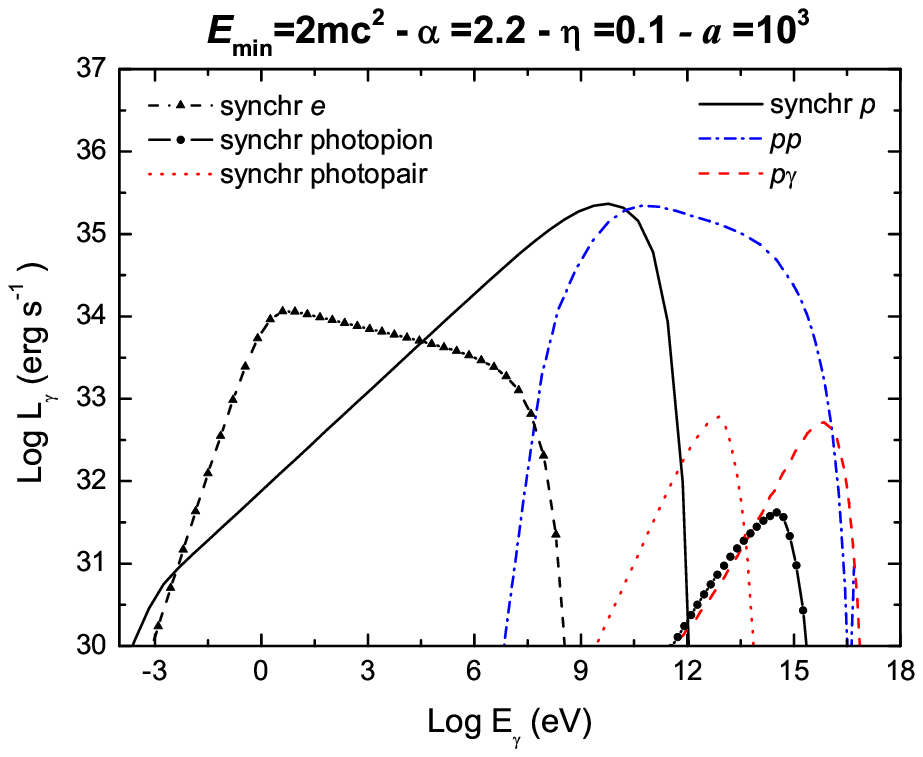} & \includegraphics[trim=10 18 10 0,clip,height=.25\textheight]{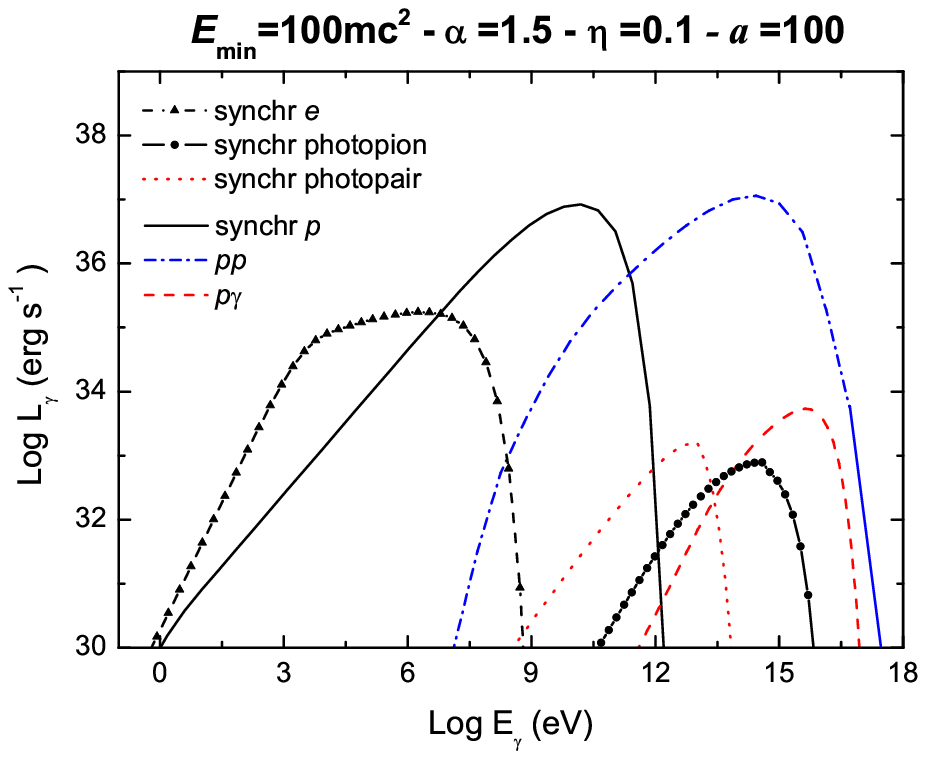} \\
\includegraphics[trim=10 18 10 0,clip,height=.25\textheight]{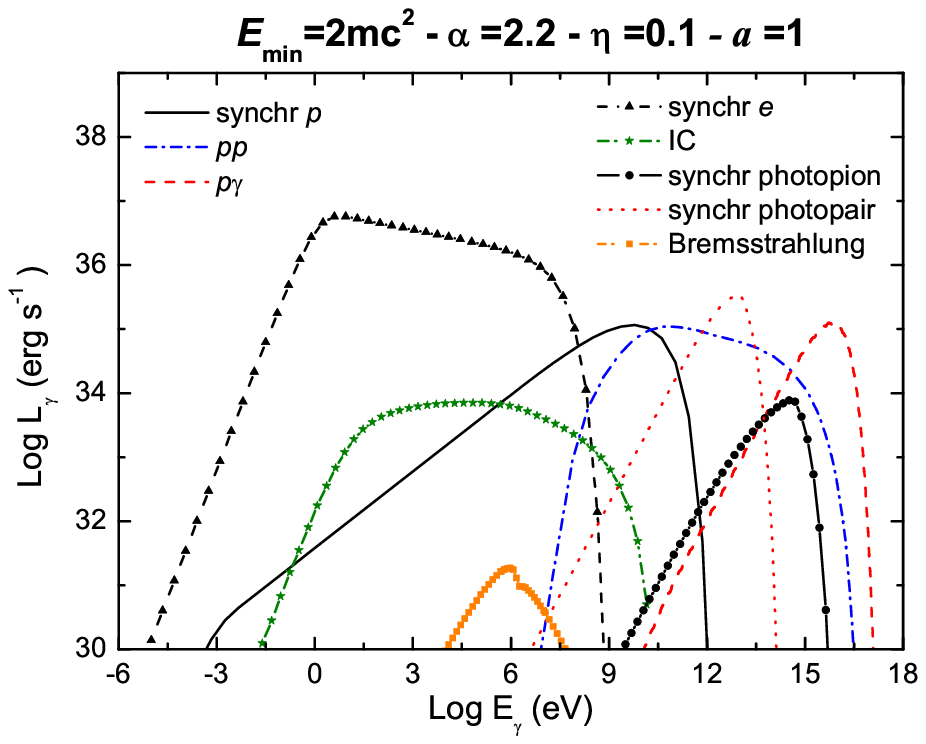}& \includegraphics[trim=10 18 10 0,clip,height=.25\textheight]{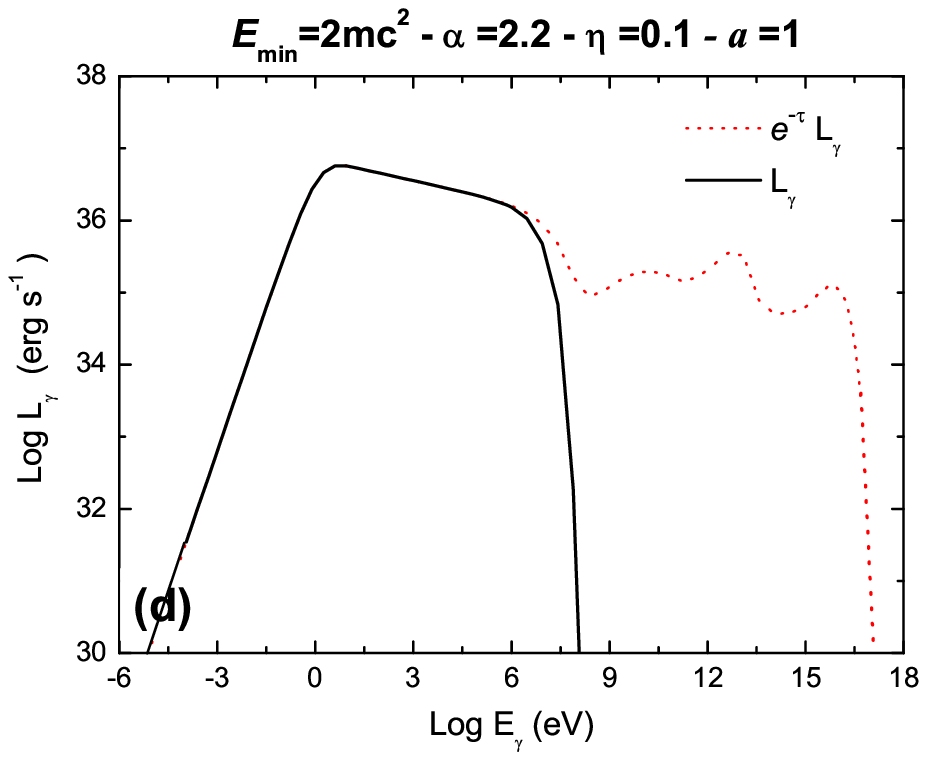} \\
\end{tabular}
  \caption{Panels (a) , (b)  and (c):  production spectra obtained for different values of the proton-to-lepton energy ratio $a$, minimum particle energy $E_{\rm{min}}$, and spectral index of the injection function $\alpha$.  Panel (d): same spectrum as in (c) corrected for absorption. }
  \label{fig:seds}
\end{figure}

Finally, we studied possible modifications of the production spectrum due to internal absorption effects: high-energy radiation in the jet can be self-absorbed by photon-photon annihilation, 

\begin{equation}
\gamma+\gamma\rightarrow e^++e^-.  
\label{neutral_pion_decay}
\end{equation}

\vspace{0.2cm}

\noindent We calculated the opacity $\tau(E_\gamma)$ for a gamma ray to escape from the emission region, and then corrected the production spectra through the attenuation factor $\exp (-\tau)$. The cross section for photon-photon annihilation and expression for the opacity can be found in \cite{Levinson2006} and \cite{GS1966}. Cascades are expected to be suppressed by the strong magnetic fields, see \cite{Khang2008}.

\section{Results}

Figure \ref{fig:seds} shows some of the spectral energy distributions (SEDs) obtained for different values of the model parameters. At energies below 1 GeV, the emission is dominated by the synchrotron radiation of  protons and primary leptons. The relative importance of the leptonic contribution depends sensibly on the parameter $a$, that fixes the proton-to-lepton total energy ratio. The electron synchrotron luminosity ranges from $10^{34} - 10^{35}$ erg s$^{-1}$ in a proton-dominated jet (cases (a) and (b)), to $10^{37}$ erg s$^{-1}$ in the case of equipartition ( $a=1$, case (c)). Case (c) is the only case where IC and Bremsstrahlung are significant, since the electron population is larger and there is an intense synchrotron photon field that serves as target for IC scattering. At energies above 1 GeV, in cases (a) and (b) the main contribution to the spectrum is due to the decay of $\pi^0$-mesons created in $pp$ collisions, reaching unabsorbed luminosities of up to $10^{37}$ erg s$^{-1}$. For $a=1$,  the synchrotron emission of secondary pairs created in $p\gamma$ interactions is relevant as well. The target field for $p\gamma$ collisions is also the synchrotron field of primary leptons, and therefore these contributions are not relevant for $a>1$. For additional details the reader is referred to \cite{RV2008}.\\
Absorption effects modify strongly the emission spectrum in the case with a strong synchrotron radiation field. In fact, as it can be seen in Figure \ref{fig:seds} (d), high-energy emission above $E_\gamma\approx1$ GeV is completely suppressed for $a=1$. In the rest of the cases, the absorption is moderate and does not result in significant changes of the emission spectrum.

\section{Conclusions}

According to the results presented in this work, low-mass microquasars could be sources of high-energy radiation. Currrent atmospheric Cherenkov telescopes are not likely to detect them, since they are sensible to photons of energy above several hundreds of GeV, and in our models emission in this energy range is completely absorbed, or the peak in the luminosity is well below the detection threshold of the detectors. However, proton microquasars could be observed by AGILE and GLAST at MeV-GeV energies, and in the future by enhanced Cherenkov arrays like MAGIC II and HESS II. Systems like those with high-levels of photomeson production should also be strong high-energy (>1 TeV) neutrino sources, since neutrinos are not affected by gamma-gamma absorption.


\begin{theacknowledgments}

We thank Matías Reynoso and Valentí Bosch-Ramon for fruitful discussions on the topics of this work. The authors were supported by the Argentine agencies CONICET (PIP 452 5375) and ANPCyT (PICT 03-13291 BID 1728/OC-AR). Additional support was provided by the Ministerio de Educación y Ciencia (Spain) under grant AYA2007-68034-C03-01, FEDER funds. G.E.R. thanks the MPIfK for kind hospitality during the preparation of this work.  

\end{theacknowledgments}

\end{document}